\documentstyle[aps,10pt,epsfig,times,latexsym,amssymb]{revtex}

\begin{document}

\title{ Implementation of Universal Control on a Decoherence-Free Qubit } 

\author{ Evan M. Fortunato${}^{1\,\ast}$, Lorenza Viola${}^{2\,\dagger}$, 
Jonathan Hodges$^{3}$, Grum Teklemariam$^{4}$, and David G. Cory${}^{1}$ }
\address{ ${}^1$ Department of Nuclear Engineering,
Massachusetts Institute of Technology, Cambridge, MA 02139 \\
${}^2$ Los Alamos National Laboratory, Mail Stop B256, Los Alamos, NM 87545 \\
${}^3$ Department of Applied Physics and Applied Mathematics, Columbia University, 
New York, NY 10027 \\
${}^4$ Department of Physics,
Massachusetts Institute of Technology, Cambridge, MA 02139 }

\maketitle

\begin{abstract}
We demonstrate storage and manipulation of one qubit encoded into a 
decoherence-free subspace (DFS) of two nuclear spins using liquid state
nuclear magnetic resonance (NMR) techniques. The DFS is spanned by states 
that are unaffected by arbitrary collective phase noise. Encoding and 
decoding procedures reversibly map an arbitrary qubit state from a single 
data spin to the DFS and back.  The implementation demonstrates the 
robustness of the DFS memory against engineered dephasing with arbitrary 
strength as well as a substantial increase in the amount of quantum 
information retained, relative to an un-encoded qubit, under both 
engineered and natural noise processes. In addition, a universal set of 
logical manipulations over the encoded qubit is also realized.  Although 
intrinsic limitations prevent maintaining full noise tolerance during 
quantum gates, we show how the use of dynamical control methods at the 
encoded level can ensure that computation is protected with finite 
distance.  We demonstrate noise-tolerant control over a DFS qubit in 
the presence of engineered phase noise significantly stronger than 
observed from natural noise sources. 
\end{abstract}



\vspace*{-1cm}
\section{Introduction}
The ability to effectively protect the coherence properties of a quantum 
information processing (QIP) device against the detrimental effects of 
environmental interactions is a prerequisite for realizing any potential 
gain of quantum computation and quantum information theory \cite{nielsen}. 
Approaches based on noiseless (or ``decoherence-free''~\cite{note}) coding 
offer a promising venue for meeting the challenge of noise-tolerant QIP. 
The theory of decoherence-free subspaces (DFSs) has been the focus of intensive 
development particularly by Zanardi, Lidar, and coworkers 
\cite{zanardi1,duanguo,lidar1,zanardipra,lidar2,lidar4,lidar3}. Recently, 
the DFS idea has been incorporated within the more general approach based on 
noiseless subsystems (NSs) \cite{viola1,viola2,qubit,zanardi3}, 
which recover DFSs and their benefits as special instances. 

The primary motivation behind ``passive'' noise control strategies relying on either
DFSs or NSs is to take advantage of specific {\sl symmetries} occurring in the noise 
process to single out subspaces or subsystems of the physical information processor 
that are inaccessible to noise. 
Once information is appropriately encoded into such noiseless structures, robust 
storage is ensured without requiring further active correction -- as long as the 
underlying symmetry dominates. These features, together with their stability against 
symmetry-perturbing errors \cite{lidar1,zanardipra,lidar4} and the consequent potential 
for concatenation with quantum error-correcting codes \cite{lidar2}, make noiseless 
codes natural candidates as robust quantum memories.  To date, experimental 
implementations include studies of DF states in quantum optical systems \cite{kwiat} ,
and one-bit quantum memories based on both a DFS of two trapped ions \cite{kielpinski} 
and a NS of three nuclear spins \cite{ns}. 

Achieving robust quantum information storage represents only a first, 
though indispensable, step toward the goal of reliable QIP. An important 
advance in this direction came from the identification of universality schemes, 
which in principle enable DFSs (or NSs) to support universal encoded quantum 
computation in a way that remains fully protected against noise. Both existential 
\cite{viola2,zanardi4} and constructive results \cite{kempe} have been established. 
While the latter are especially appealing for a class of proposed quantum computing 
architectures governed by Heisenberg exchange interactions \cite{exchange}, 
implementations of these schemes remain difficult due to the stringent symmetry 
and tunability requirements on the control Hamiltonians.

Here, we take a first experimental step towards encoded quantum computation by 
demonstrating universal control over a one-bit DF quantum register of two nuclear 
spins. A novel key ingredient we use to implement encoded quantum gates is the 
combination of robust control design with the use of dynamical decoupling methods 
\cite{CP,cory,viola3,viola4} directly on encoded degrees of freedom.  
Our results suggest that this may serve as a useful strategy for practically 
coping with the constraints required for DF computation.

The paper is organized as follows. In Sect. II we review the collective decoherence
model that is relevant to the work, along with the prescriptions from the DFS theory
for both protected storage and manipulation of quantum information in a two-qubit 
system. In Sect. III, we outline our proposed approach to noise-tolerant control
of DFS encoded qubits based on {\sl concatenating encoded decoupling methods with 
robust control design}. The general principles are developed starting from the 
physical NMR setting relevant to the experiment. Sect. IV contains an account of 
the control techniques used in the experiment and the reliability measures 
adopted to quantify the accuracy of the implementation. In particular, a notion
of {\sl gate entanglement fidelity}, generalizing Schumacher's definition to 
allow a desired unitary evolution on the quantum data, is proposed and related 
to other fidelity metrics relevant to QIP. The experimental results demonstrating 
protected storage and universal protected quantum logic are presented and 
discussed in Sect. V and VI, respectively.

\section{Protecting Quantum Information against Collective Decoherence}
\subsection{Collective decoherence}

For a system $S$ composed of $n$ qubits, a purely decohering, collective interaction 
arises when the qubits couple symmetrically to a single environment $E$ and no exchange
of energy takes place between $S$ and $E$. Physically, this model accounts for 
relaxation due to {\sl fully correlated fluctuations} of the energy levels of each qubit 
-- a situation that is approached if the qubits are close enough relative to the 
correlation length of the environmental coupling and the latter {\sl commutes} with 
the natural Hamiltonian. 
Although not always applicable, this decoherence model has a practical significance 
for QIP. In particular, collective dephasing was shown to play a major role
in quantum devices based on trapped ions \cite{kielpinski}. In NMR systems, 
dephasing caused by fully correlated fluctuations of the local magnetic field
provides the dominant relaxation mechanism of quantum coherences between
identical species in sufficiently small, rigid molecules \cite{ernst}. 

If $H = H_S \otimes \openone_E + \openone_S \otimes H_E + H_{SE}$ represents 
the Hamiltonian for the joint system plus environment, collective phase damping 
corresponds to an interaction Hamiltonian of the form 
\begin{equation}
H_{SE} = J_z \otimes B_z \:, 
\label{intham}
\end{equation}
where the operator $B_z$ acts only on $E$, and $J_z$ measures (in units $\hbar/2$)
the projection of the total spin angular momentum along the quantizing axis $\hat{z}$.
Thus, $J_z = \sum_{j=1}^n \sigma_z^j$ in terms of the Pauli operator $\sigma_z^j$
acting on the $j$th qubit. Starting from an initial state 
$\varrho_{in}=|\Psi_{in}\rangle\langle \Psi_{in}|$, the system alone
evolves after a time $t$ into $\varrho_{in} \mapsto \varrho_{out}= 
{\cal S}(\varrho_{in})$, where ${\cal S}$ is the super-operator associated with $H$ 
\cite{nielsen}. 
Under the assumption that $S$ and $E$ are initially uncorrelated, $\varrho_{out}$ 
can be expressed as 
\begin{eqnarray*}
\varrho_{out}= \sum_a F_a \varrho_{in} F_a^\dagger\:, \hspace{1cm} 
\sum_a F_a^\dagger F_a =\openone_S \:, 
\end{eqnarray*}
for a set of Kraus operators $\{F_a\}$ \cite{kraus}. Because $[H_S, J_z]=0$ for pure
decoherence, the unitary contribution due to $H_S$ can be separated out as 
$F_a=E_a \exp(-iH_St)$, leaving a set of error operators $\{ E_a\}$,
that still satisfy $\sum_a E_a^\dagger E_a =\openone_S$, and describe the 
non-unitary effects of the environment. 
If $t$ is sufficiently short (or, equivalently, the coupling is weak enough),
the operators $E_a$ can be expressed as linear combinations of the basic error
generator $J_z$ and the identity. For arbitrary collective decoherence, the possible 
errors that the coupling (\ref{intham}) can induce belong to the {\sl interaction
algebra} \cite{viola1} ${\cal A}_z$ generated by $J_z$ {\it i.e.}, the 
algebra containing all the linear combinations of arbitrary 
powers of $J_z$ and $\openone_S$. By construction, ${\cal A}_z$ is Abelian, 
expressing the fact that pure decoherence is energy-conserving -- thus diagonal 
in the computational basis.
 
We focus on $n=2$ qubits, in which case every element in ${\cal A}_z$ 
can be written as a linear combination of three operators, 
$\openone, J_z=\sigma_z^1+\sigma_z^2, J_z^2$ -- equivalently, we replace the latter
with $\sigma_z^1 \sigma_z^2=J_z^2/2 -\openone$ ($\openone =\openone_S$ henceforth). 
If $\{ |00\rangle, |01\rangle, |10\rangle, |11\rangle\}$ denotes the 
computational basis, an equivalent choice as a basis for ${\cal A}_z$ are the 
three orthogonal projectors on subspaces with definite $\hat{z}$-angular momentum, 
$\Pi_{+2}=|00\rangle \langle 00| = (\openone+J_z+\sigma_z^1\sigma_z^2)/4$, 
$\Pi_{0}=|01\rangle \langle 01| + |10\rangle \langle 10|= 
(\openone-\sigma_z^1\sigma_z^2)/2$, $\Pi_{-2}=|11\rangle \langle 11|=
(\openone-J_z+\sigma_z^1\sigma_z^2)/4$. In addition to labelling the basis states, 
the $J_z$ quantum number also allows classification of the possible 
transitions between these levels, via the so-called {\sl coherence order} 
\cite{ernst}. For a quantum coherence between two states with $\hat{z}$-angular 
momentum $k, \ell$, corresponding to the off-diagonal density matrix
element $|k\rangle \langle \ell|$
($k,\ell$ being now measured in units $\hbar$), the coherence order
$m_{k \ell}=|k-\ell|$. The behavior of the zero-, one-, and two-quantum coherence 
orders present in a two-spin system undergoing collective decoherence can be described 
by obtaining an explicit set of error operators $\{E_a\}$. This is done starting 
from a unitary representation $U_{SE}$ for the joint $SE$ evolution \cite{kraus},
\begin{eqnarray*}
        & & |00\rangle |e\rangle \rightarrow |00\rangle |e_0\rangle \:, \\
        & & |01\rangle |e\rangle \rightarrow |01\rangle |e_1\rangle \:, \\
        & & |10\rangle |e\rangle \rightarrow |10\rangle |e_1^\prime\rangle \:, \\
        & & |11\rangle |e\rangle \rightarrow |11\rangle |e_2\rangle \:,
\end{eqnarray*}
where the $|e\rangle, |e_k\rangle$ are generally non-orthogonal environment states,
and the collective nature of the interaction sets $\langle e_1| e_1^\prime\rangle =1$. 
The overlaps $\langle e_0|e_1\rangle=\langle e_1|e_2\rangle =e^{-\gamma}$, 
$\langle e_0| e_2\rangle =e^{-\gamma^\prime}$ parametrizes the decay rates for 
single and double quantum coherences, respectively. 
By letting $\gamma^\prime = 4 \gamma$ ($\gamma \geq 0$) \cite{ernst}, and by 
using an orthonormal basis $|\mu_a\rangle$ obtained from the $|e_a\rangle$, 
$a=0,1,2$, a choice of Kraus operators is given by 
$E_a = \langle \mu_a | U_{SE} | e\rangle$ {\it i.e.},
\begin{eqnarray}
\left\{ \begin{array}{lll}
E_0 & = &\Pi_{+2} + e^{-\gamma} \Pi_0 + e^{-4 \gamma} \Pi_{-2} \:, 
\\
E_1 & = & \sqrt{1-e^{-2\gamma} } \,\Pi_0 + e^{-\gamma} (1+e^{-2\gamma}) 
\sqrt{1-e^{-2\gamma}} \,\Pi_{-2} \:, 
\\
E_2 & =& (1-e^{-2\gamma}) \sqrt{1+e^{-2\gamma}} \, \Pi_{-2} \:. 
\end{array} \right. 
\label{kraus}
\end{eqnarray}
An equivalent representation was derived for dephasing quantum operations 
engineered via gradient-diffusion techniques in NMR \cite{havel}. In the limit
of arbitrarily strong (or ``crusher'') interaction, $\gamma \rightarrow \infty$, 
Eqs. (\ref{kraus}) simplify to $E_0=\Pi_{+2}, E_1=\Pi_0, E_2=\Pi_{-2}$,
leading to the full suppression of single and double quantum coherences. However, 
regardless how strong, collective decoherence produces {\sl no decay of the} 
{\sl zero-quantum subspace} spanned by $\{ |01\rangle,|10\rangle\}$.

Zero-quantum coherences and their properties have been long appreciated in 
NMR, with important applications in both high-resolution spectroscopy in 
inhomogeneous magnetic fields and contrast enhancement in magnetic 
imaging \cite{hall,warren}.  Within NMR QIP, zero-quantum coherences are
revisited in view of their natural potential to {\sl encode} protected 
quantum information.

\subsection{Decoherence-free encodings}

In the DFS approach, the first step is ensuring that the quantum data to be 
protected is encoded into a DFS.
Mathematically, a DFS is a subspace of the system's state space spanned by a 
set of {\sl degenerate eigenvectors} of all the error generators appearing in 
$H_{SE}.$ For global dephasing on $n$ qubits as in (\ref{intham}), let 
${\cal H}^{(j_z)}$ be the eigenspace corresponding to the eigenvalue $j_z$ of $J_z$, 
$j_z=n, n-2,\ldots,-n+2,-n$.
Then states in ${\cal H}^{(j_z)}$ remain {\sl invariant} under the 
environmental coupling, 
\begin{equation}
J_z |\psi_L\rangle = j_z |\psi_L\rangle \:, \hspace{1cm} \forall |\psi_L \rangle \in 
{\cal H}^{(j_z)} \:,
\label{inv1}
\end{equation}
which also implies a degenerate action of {each} error operator on the subspace:
\begin{equation}
E_a |\psi_L\rangle = f_a |\psi_L\rangle \:, \hspace{1cm} \forall |\psi_L \rangle \in 
{\cal H}^{(j_z)},\;\; \forall E_a \in {\cal A}_z \:,
\label{inv2}
\end{equation}
for some coefficients $f_a = f_a(j_z)$ fulfilling $\sum_a |f_a|^2=1$. If the system
is initialized in a state 
$\varrho_{in}=|\psi_L\rangle\langle \psi_L| \in {\cal H}^{(j_z)}$, then
\begin{eqnarray*}
\varrho_{out}= \sum_a E_a e^{-i H_S t} |\psi_L\rangle\langle\psi_L| e^{+i H_S t}
E_a^\dagger = \Big( \sum_a |f_a|^2 \Big) e^{-i H_S t} |\psi_L\rangle\langle\psi_L| 
e^{+i H_S t} = e^{-i H_S t} \varrho_{in} e^{+i H_S t}  \:, 
\end{eqnarray*}
{\it i.e.}, the evolution remains unitary within each ${\cal H}^{(j_z)}$. Thus, each 
${\cal H}^{(j_z)}$ is a DFS under collective decoherence. The amount of quantum 
information that a given DFS is able to protect is determined by its 
dimension $n_{j_z}$ -- which is simply the degeneracy of the corresponding
$j_z$-eigenvalue. In particular, for $n$ even, the largest DFS is supported by the 
zero-quantum subspace ${\cal H}^{(0)}$, with $n_0= n!/(n/2!)^2$. 

For $n=2$ spins, ${\cal H}^{(0)}$ is doubly degenerate, hence it provides the 
smallest DFS capable to protect one qubit against collective decoherence. 
The robustness of this two-spin zero-quantum subspace  under ${\cal A}_z$ has 
been explicitly derived above.
In terms of basis states, our DFS qubit is defined by the encoding
\begin{equation}
c_0 |0_L\rangle + c_1 |1_L\rangle =
c_0 |01 \rangle + c_1 |10\rangle \:, \hspace{1cm}
{\cal H}_L={\cal H}^{(0)}= \text{span}\{ |0_L\rangle , |1_L\rangle \}\:,
\label{qubit}
\end{equation}
for arbitrary complex coefficients $c_0, c_1$. 
It is worth stressing that the existence of a DFS is tied, at the physical 
level, to the occurrence of {\sl symmetries} in the noise process. 
The way the latter reflect into the state space of the system both
determines the possibility of invariant states as in (\ref{inv1})-(\ref{inv2})
and the associated degeneracies. For any interaction which is diagonal 
in the computational basis, the underlying ``axial'' symmetry ensures that the 
individual $\sigma^j_z$ are conserved quantum numbers. However, it is only under 
the additional permutation symmetry characterizing collective interactions that the 
{\sl degenerate} conserved quantum number $J_z$ arises -- signaling the presence of 
a protected structure.  The emergence of degenerate degrees of freedom preserved 
under the noise remains the key ingredient for more general DFSs 
\cite{zanardi1,lidar1,kempe} and, in still more elaborated forms, NSs as well 
\cite{viola1,qubit,ns}. Taking advantage of the existing symmetries translates
into major gains toward achieving noise-protected QIP. For instance, the simple 
2-bit encoding (\ref{qubit}) preserves a qubit against collective dephasing of 
arbitrary strength, to be contrasted with independent phase errors -- where 
protection can be achieved only with finite distance using a quantum 
error-correcting code.

\subsection{Decoherence-free manipulations} 

Once protected storage of quantum information is obtained, the next step is to 
ensure that universal quantum gates are implemented without ever leaving the DFS. 
Because the states spanning a DFS are characterized by precise symmetry properties, 
symmetries are likewise crucial in determining the control operations to be applied 
for effecting DF quantum logic. Clearly, the allowed gates must map DFS states to 
DFS states. However, as any physical gate takes a finite time to execute, invoking 
unitary manipulations that preserve the DFS at the conclusion of the gate is not 
sufficient. To guarantee that the system remains within the DFS during the entire 
gating time requires the stronger condition that gates are generated by Hamiltonians 
that themselves respect the symmetry. Thus, the general problem requires identifying 
a universal set of Hamiltonians which satisfy the correct symmetry constraints and 
involve at most two-body interactions \cite{kempe}. 

In our case, because a single DFS qubit is involved, this universal set of control 
Hamiltonians is composed of two observables generating an encoded u(2) Lie 
algebra, exponentiation then giving the whole group U(2) of encoded one-qubit 
transformations. A necessary and sufficient condition for an Hamiltonian $A$ 
to preserve the zero-quantum DFS can be obtained by demanding that 
$\langle 00| A (c_0 |01\rangle + c_1 |10\rangle) = 0$, 
$\langle 11| A (c_0 |01\rangle + c_1 |10\rangle) = 0$ for arbitrary $c_0,c_1$.
This leads to the following matrix form for $A$ with respect to the computational 
basis:
\begin{eqnarray}
A= \left( \begin{array}{cccc}
       a_1 & 0   & 0 & c \\
       0   & a_2 & b & 0 \\
       0   & b^* & a_3 & 0 \\
       c^* & 0 & 0 & a_4 
\label{goodham}
\end{array}\right)
\end{eqnarray} 
for real coefficients $a_j,\,j=1,\ldots,4$, and possibly complex $b,c$. In 
particular, this includes all the hermitian operators belonging to the so-called
{\sl commutant} of the error algebra \cite{viola1,kempe}, 
${\cal A}'_z=\{X: [X,J_z]=0\}$, which collects all operators commuting with 
the noise. Because every operator in ${\cal A}'_z$ can be represented as a linear 
combination of the identity,  the one-bit operators $\sigma_z^j$, and the two-bit 
couplings $\sigma_z^1\sigma_z^2$, $\sigma^1\cdot \sigma^2=\sigma_x^1\sigma_x^2
+ \sigma_y^1\sigma_y^2 + \sigma_z^1\sigma_z^2$ (Heisenberg coupling), Hamiltonians 
in ${\cal A}'_z$ have $c=0$ -- implying that {\sl all} the DFSs are in fact preserved. 
An additional constraint (so-called {\sl independence} \cite{kempe}) can be imposed 
on Hamiltonians in ${\cal A}'_z$ by also requiring $a_1=a_4=0$, in which case 
$A$ has zero entries outside the selected zero-quantum  DFS. 

With respect to the DF encoding (\ref{qubit}), a choice of operators that 
act as independent, encoded $\sigma_z$, $\sigma_x$ on ${\cal H}_L$ 
is given by
\begin{eqnarray}
\left\{ \begin{array}{lll}
\sigma_z^L &=_L & {1 \over 2} \Big( \sigma_z^1-\sigma_z^2 \Big)\:,
\\
\sigma_x^L &=_L & {1\over 2} \Big(\sigma^1\cdot \sigma^2 - \sigma_z^1 \sigma^2_z \Big)=
        {1\over 2} \Big(\sigma^1_x\sigma^2_x + \sigma_y^1 \sigma^2_y \Big)\:,
\end{array} \right. 
\label{obs1}
\end{eqnarray}
where the notation $=_L$ means equality upon restriction to ${\cal H}_L$ 
and $\sigma_y^L=_L i[\sigma_x^L,\sigma_z^L]/2$.  
The independence property is useful to allow parallel encoded manipulations 
on different DFSs. However, when only a single DFS is in use, requiring 
independence or even preservation of all DFSs has no advantages, and allowing 
for the most general Hamiltonian as in 
(\ref{goodham}) may in fact increase the options available for implementation.
For instance, an alternative choice for encoded $z$ and $x$ observables is
\begin{eqnarray}
\left\{ \begin{array}{lll}
\sigma_z^L &=_L & - \sigma_z^2  \:, 
\\
\sigma_x^L &=_L & \sigma_x^1 \sigma_x^2 \:, 
\end{array} \right. 
\label{obs2}
\end{eqnarray}
and again $\sigma_y^L=_L i[\sigma_x^L,\sigma_z^L]/2$. 

In principle, based on standard universality results \cite{nielsen}, it is possible to 
generate any encoded unitary transformation by appropriately alternating evolutions 
under {\sl two} Hamiltonians with the correct symmetry. For instance, this is certainly
true if one can turn on/off a pair of Hamiltonians in ${\cal A}'_z$ such as, say, 
$\sigma_z^1$ and the exchange interaction $E_{12}=(\sigma^1\cdot \sigma^2 + 
\openone)/2$ -- for $i[E_{12},\sigma^1_z]/2$ gives an encoded $\sigma_y^L$ and then 
$-i[E_{12},\sigma_y^L]/2$ gives $\sigma_z^L$ as in (\ref{obs1}). 

Once encoded single-qubit manipulations are available, then universal encoded
computation over DFS qubits requires the additional ability of implementing
a non-trivial encoded gate between two logical qubits. For instance, a 
controlled-rotation gate could be constructed from a logical phase 
coupling of the form \mbox{$\sigma_z^{L_j}\sigma_z^{L_{j'}}$}, which is supported by
the natural couplings of many NMR and NMR-like Hamiltonians. We focus here on
the first step of this program {\it i.e.}, to obtain reliable single-qubit
DF manipulations compatible with the constraints that QIP implementations 
unavoidably face in terms of both the form and the tunability of the available 
control Hamiltonians.

\section{Controlling encoded quantum information with reduced error rate}

\subsection{Two-spin NMR QIP as a case study}

The total system Hamiltonian we consider, $H_S$, is the sum of a time-independent 
internal Hamiltonian, $H_{int}$, and a time-dependent external 
Hamiltonian, $H_{ext}$.  The internal Hamiltonian, composed of
spin-field and spin-spin interactions, is \cite{cory}
\begin{eqnarray}
H_{int}=\pi(\nu_1\sigma_z^1 + \nu_2\sigma_z^2 + J\sigma^1\cdot\sigma^2/2)\:,
\label{hint}
\end{eqnarray}
where $\nu_1, \nu_2$, and $J$ are the chemical shifts and the coupling 
constant, respectively.  The external Hamiltonian, describing the 
interaction between the spins and an applied RF field has the form 
\cite{cory,control}
\begin{equation}
H_{ext}= \sum_{k=1,2} 
e^{-i(\omega_{RF} t+\phi) \sigma_z^k/2 } 
(\omega \sigma^k_x/2) e^{i(\omega_{RF}t+\phi)\sigma^k_z/2 } \:,
\label{hext}
\end{equation}
the transmitter's angular frequency $\omega_{RF}$, the initial phase $\phi$,
and the power $\omega$ being tunable over an appropriate parameter range.  

The implementation of an arbitrary unitary gate is accomplished by 
modulating $H_{int}$ via an external control sequence.  While sequences
can be optimized numerically \cite{control}, average Hamiltonian theory 
(AHT) \cite{Waugh} provides a systematic method for describing any
unitary propagator $U(T)$ resulting from the evolution under the time-varying
Hamiltonian $H_S=H_{int}+H_{ext}$ in terms of an effective Hamiltonian 
$\overline{H}$ applied 
over the same time interval:
\begin{eqnarray*}
U(T)= {\cal T}\hspace*{-1mm}\exp \Big(\hspace*{-1mm} -i \int_0^T d\tau 
H_S  (\tau) \Big) 
= e^{-i \overline{H} T} \:,
\end{eqnarray*}
where ${\cal T}$ is, as usual, the Dyson time-ordering symbol. In particular,
AHT underlies the design of coherent {\sl refocusing} and {\sl decoupling} 
methods, which are able to effectively turn on/off selected contributions to the 
average propagator over some time interval. These methods have been recently revisited 
within the QIP context in \cite{viola3,viola4,viola2}. We recall that the basic idea
is to subject the system to a cyclic train of pulses 
${\cal P} = \{ P_j\}_{j=1}^M$, $\Pi_{j=1}^M P_j =\openone$ which, in the simplest 
setting, are assumed to be infinitely short and equally spaced by $\Delta t >0$.
The net controlled evolution over the period $T=M\Delta t$ can then be 
expressed as
\begin{eqnarray*}
e^{-i \overline{H} T} = \prod_{k=0}^M e^{-i H_k \Delta t}  \:,
\end{eqnarray*} 
where the ``toggling-frame'' Hamiltonians $H_k$ are determined as 
$H_k=U_k^\dagger H_{int} U_k$, in terms of the composite pulses 
$U_k=\Pi_{j=1}^k P_j$, $k=1,\ldots,M$, $U_0=\openone$ \cite{ernst}. 
In the limit of sufficiently rapid control, 
$\overline{H}$ simply approaches \cite{ernst,viola3}
\begin{eqnarray*}
\overline{H} = {1 \over M} \sum_{k=0}^M H_k = {1 \over M} \sum_{k=0}^M
U_k^\dagger H_{int} U_k \:.
\end{eqnarray*}
By appropriately designing the pulse sequence ${\cal P}$, undesired 
contributions to $\overline{H}$ can be effectively turned off. 
For instance, a train  ${\cal P}_1$ of equally spaced, simultaneous $\pi_x$ 
pulses on both spins ($\pi^1_x\pi^2_x$ pulses) 
averages out any phase evolution due to the $\sigma_z^j$ terms in 
(\ref{hint}). Similarly, the $\sigma_x^1\sigma_x^2 + \sigma_y^1\sigma_y^2$ 
coupling can also be averaged to zero by a pulse sequence ${\cal P}_2$ 
consisting of repeated, equally spaced $\pi^1_x\pi^2_y$ pulses.

\subsection{Universal gates via encoded dynamical control}

While AHT represents a powerful tool for designing logic gates over physical, 
{\sl un-encoded} degrees of freedom, a direct application on DF encoded qubits 
does not automatically result in DF manipulations. 
Even though ensuring that $\overline{H}$ has the general form (\ref{goodham}) 
(for instance, $\overline{H} \in {\cal A}'_z$) leaves the system in a DF state, 
there is no guarantee that the control path has remained within the DFS at all
intermediate times -- possibly re-introducing exposure to noise. 

We begin by noting that $H_{int}$ can be rewritten as 
\begin{equation}
H_{int}= \pi \Big( {\nu_1+\nu_2 \over 2} J_z + {J \over 2} \sigma_z^1 \sigma_z^2
+ (\nu_1-\nu_2) \sigma_z^L + J \sigma_x^L \Big) \:, 
\label{hint2} 
\end{equation}
in terms, for instance, of the encoded observables (\ref{obs1}) -- which makes it 
explicit that $H_{int} \in {\cal A}'_z$. Since both $J_z$ and $\sigma_z^1 \sigma_z^2$ 
are constant on the code subspace ${\cal H}_L$, they can be ignored and $H_{int}$ 
further simplifies to
\begin{equation}
H_{int} =_L \pi ( \Delta \nu \,\sigma_z^L + J \,\sigma_x^L ) \:, 
\hspace{1cm} \Delta \nu = \nu_1-\nu_2 \:.
\label{hint3} 
\end{equation}
Thus, the natural evolution implements a non-trivial logical operation 
within ${\cal H}_L$. The challenge is to extract the required controlled 
operations by remaining, ideally, always within the DFS.

The situation is simpler in the limit where, as above, 
control pulses are treated as instantaneous.  Because $H_{int} \in {\cal A}'_z$, 
one can ensure that each toggling-frame Hamiltonian $H_k$ also remains in ${\cal A}'_z$
by choosing pulses such that either $[U_k,J_z]=0$ or $\{ U_k,J_z\}=0$. The latter
condition is satisfied, for instance, by the above-mentioned  pulse sequence 
${\cal P}_2$, which thus implements a net encoded identity in this idealized scenario.  

Of course, the duration of real-life pulses is necessarily finite,  
and one needs to pay additional care to what happens during the pulse length 
\cite{haeberlen}. 
In principle, DF logical operations can still be effected if sufficient control over 
the parameters $\Delta \nu$, $J$ in (\ref{hint3}) is available. 
The general idea is to concatenate AHT with the underlying DF encoding 
{\it i.e.}, to {\sl implement refocusing directly with encoded rotations}.  
Let us look at our DFS qubit (a more expanded account will be provided elsewhere; 
see also \cite{wu} for related work), and imagine that encoded $\pi^L$ pulses are 
available as $\pi_{x,y}^L=\exp(-i\pi \sigma^L_{x,y}/2)$.
Then a sequence of equally spaced encoded $\pi^L_x$ pulses (in this case a 
Carr-Purcell sequence~\cite{CP}) can be used to refocus the encoded phase 
evolution and only leave the encoded $\sigma_x^L$ coupling active in (\ref{hint3}). 
This can be thought of as a logical or encoded ``spin echo" \cite{Hahn}.  
A similar procedure holds for extracting the encoded $\sigma_z^L$ Hamiltonian 
if encoded $\pi^L_z$ pulses are employed instead.  
Thus, the same schemes that are effective at turning on/off unwanted 
terms in the physical qubit evolution are effective at turning on/off unwanted terms 
in the encoded qubit evolution, provided ordinary control pulses are replaced with 
encoded ones.
More generally, a group-theoretical framework extending the un-encoded approach
of \cite{viola3} to {\sl encoded dynamical decoupling} can be constructed. 
For our system, this implies that the ability to apply a {\sl single} Hamiltonian
with the correct symmetry ({\it e.g.}, $\sigma_x^L$) suffices, in principle, for
gaining universal control.

Unfortunately, such control is not directly available in practice, as the 
evolutions induced by the external RF Hamiltonian (\ref{hext}) do not resemble, 
in general, evolutions under logical Hamiltonians. The approach we take results 
from the following compromise: we {\sl mimic} the implementation of a fully
encoded refocusing scheme by using available pulses whose {\sl propagator}
(not Hamiltonian) equals the required encoded rotation; we then 
compensate for the residual exposure to noise by control design. 
If pulse durations are optimized, then the system will reside in the 
DFS for a dominant portion of the computational time.  In addition, pulse 
design can add robustness against noise \cite{Shaka1}, reducing its impact 
while the system resides outside the protected space.  While in the limit of 
weak noise with arbitrarily long correlation times these techniques provide 
robustness, for realistic noise models actual improvements will depend heavily 
on the noise parameters. An explicit implementation will be reported.

As already noted, these ideas open the way for manipulating more than a single
encoded qubit.  If, for instance, two DFS qubits are supported by 
the zero-quantum 
subspaces of, say, two proton and two carbon spins, the overall internal Hamiltonian 
will be expressible, to high accuracy and for a wide class of spin-spin coupling 
distributions, in terms of both single-qubit encoded observables $\sigma_x^{L_{1,2}}$, 
$\sigma_z^{L_{1,2}}$ and the two-qubit encoded interaction 
$\sigma_z^{L_1}\sigma_z^{L_2}$. Thus, the ability of separately controlling each 
encoded qubit via encoded refocusing, combined with the presence of the logical 
phase coupling, implies the potential of effecting universal quantum logic 
with reduced error rate.

\section{Experimental outline}

Liquid state NMR QIP techniques have been extensively discussed in the 
literature \cite{cory}, and only the salient points are recalled here.  
Because the system exist in highly mixed, separable states, NMR QIP relies 
on ``pseudo-pure'' (p.p.) states whose traceless, or deviation, component 
is proportional to that of the corresponding pure state.  The identity 
component of the density matrix is unobservable and is treated as a constant 
under the assumption of unital dynamics ({\it i.e.}, dynamics that preserves 
the completely mixed state). In this case, the evolution of a p.p. state is 
equivalent to the corresponding pure-state evolution. Initialization of the 
two-spin system into an intended p.p. state was accomplished using 
gradient-pulse techniques as described in \cite{cory,sodickson}.  Throughout 
the experimental implementation, all deviation components were explicitly 
verified by state tomography \cite{Tomo}. A fixed amount of identity 
component that optimizes the fidelity between the experimentally determined 
and a desired reference p.p. state, $|00 \rangle \langle 00|$, was added 
to each reconstructed deviation density matrix. For each experiment, we 
prepared one of the p.p. input states 
$\varrho^{p.p.}_{in} = | \psi_{in} 0 \rangle \langle \psi_{in} 0 |$,
with $|\psi_{in} \rangle \langle \psi_{in} |$ providing a complete set 
of one-bit density matrices so as to allow quantum process tomography 
reconstruction \cite{ike,ns}.    

Our physical system is an ensemble of Dibromothiophene molecules (Fig. 1)
in a solution of CDCl$_3$.  Measured values for the relevant parameters are 
listed in the caption of Fig. 1. The experimental procedure begins with the data 
qubit 1 containing the state $|\psi_{in}\rangle $ to be protected, 
$|\psi_{in}\rangle = c_0 |0\rangle + c_1 |1\rangle$, and the ancilla 
qubit 2 initialized to $|0\rangle$. Encoding of the initial input state 
to the code space ${\cal H}_L$ is accomplished by the unitary 
transformation 
\begin{eqnarray*}
U_{enc} (c_0 |0\rangle + c_1 |1\rangle)_1 |0\rangle_2 =
c_0 |0_L\rangle + c_1 |1_L\rangle \:, 
\end{eqnarray*}
where $U_{enc}$ is a controlled $\sigma_x$ rotation on bit 2 conditioned on 
bit 1 having the state $|0\rangle$. 
Next, an encoded operation is performed on the system in the presence of noise.
The information is retrieved by applying a decoding transformation 
$U_{dec}=U_{enc}^\dagger$, producing a general output state of the form
\begin{equation}
\varrho_{out}= U_{target}|\psi_{in}0\rangle \langle \psi_{in}0| 
U_{target}^\dagger = U_{target}|\psi_{in}\rangle_1 \langle \psi_{in}| 
U_{target}^\dagger \otimes |0\rangle_2\langle 0| \:,
\label{out} 
\end{equation}
for a target single-qubit unitary transformation $U_{target}$ on the data spin.
$U_{target} =\openone$ corresponds to storage of the quantum data under 
either engineered collective dephasing or natural noise, while $U_{target}$
is a non-trivial desired rotation for demonstrating universal quantum logic.
All experiments were carried out on a 400 MHz Bruker {\sc avance} spectrometer.

\subsection{Unitary and non-unitary control}

The un-encoded gate operations involved in the encoding and decoding 
networks were mapped into ideal pulse sequences using standard methods 
\cite{cory}.  Pulses were then implemented by modulating $H_{int}$ with 
external RF fields as mentioned in Sect. IIIa \cite{cory}.  

Non-unitary evolution, either for p.p. state preparation or for emulating 
collective decoherence, were implemented using pulsed magnetic field gradients.  
Magnetic field gradients take advantage of the spatial extent of the sample to 
induce an incoherent evolution.  Applying a gradient $\nabla_z B= \partial 
B_z / \partial z$ along the axis of the static field causes a linear variation of 
the Larmor precession frequency given by the spatially dependent Hamiltonian  
\begin{eqnarray*}
H_{grad} = \gamma z J_z \nabla_z B  /2 \:,
\end{eqnarray*}
$\gamma$ being the gyro-magnetic ratio of the given nuclear species. 
This causes each quantum coherence $\rho_{k\ell}$ ($k\ne\ell$) to be multiplied 
by a spatially dependent phase factor, 
$\exp(-i \gamma z m_{k\ell} \nabla B_z  \delta / 2)$, where $m_{k\ell}$ is 
the coherence order (defined earlier) and $\delta$ the duration of the gradient 
pulse. In other words, each part of the sample experiences a different 
coherent phase error.  Tracing over the spatial degrees of freedom, 
as is done in an ensemble measurement, causes this incoherent evolution 
to become irreversible when considering the spin degrees of freedom alone.  
While the effects of this evolution could be immediately reversed, random 
molecular diffusion causes an irreversible spatial displacement that increases 
with both time and the molecular diffusion coefficient.  Applying an 
inverse gradient after a time delay $\Delta$ (diffusion time) thus results in 
an exponential decay of non-zero coherences, $\exp(- \Delta/\tau)$, with an 
effective noise strength given by \cite{sodickson} 
\begin{eqnarray*}
{1 \over \tau} = D (\gamma \nabla_z B \, m_{k\ell}\delta)^2 \:,
\end{eqnarray*}
$D$ being the diffusion coefficient of the sample. Note the scaling of this 
decoherence rate with the {\sl square} of the coherence order, as anticipated in 
the derivation of Eqs. (\ref{kraus}).

Using these gradient-diffusion techniques, variable strength noise can be 
obtained by either changing the gradient strength or the diffusion time.  
It should be noted that, in both the incoherent and decoherent case, the induced 
phase error is collective to an extremely good extent, deviations from a collective 
action being determined by the product of the gradient strength 
(approximately 60 Gauss/cm) and the spatial displacement between the two 
hydrogen spins (on the order of angstroms).

\subsection{Reliability measures for control}

As a reliability measure quantifying the accuracy of implementing a target
unitary transformation $U$ on a system $S$ we invoke a variant of the 
entanglement fidelity $F_e$ as introduced by Schumacher \cite{schumacher}.  
In Schumacher's notation, let $R$ be an auxiliary ``reference'' system, and 
let the initial entangled state $|\Psi^{RS}\rangle$ of the pair $RS$ be 
subjected to the overall evolution $\openone^R \otimes {\cal E}^S$. 
Starting from $\rho^{RS}=|\Psi^{RS}\rangle \langle \Psi^{RS}|$, 
this produces a final state 
$\rho^{RS'} = (\openone^R \otimes {\cal E}^S ) (|\Psi^{RS}\rangle 
\langle \Psi^{RS}|)$. Then the entanglement fidelity of the process 
${\cal E}^S$ relative to the initial state of $S$ alone, 
$\rho^S= \mbox{Tr}_R \{ |\Psi^{RS}\rangle \langle \Psi^{RS}|\}$, is defined as
\begin{equation}
F_e(\rho^S, {\cal E}^S)= \mbox{Tr} \Big\{ |\Psi^{RS}\rangle \langle \Psi^{RS}|
\, \rho^{RS'} \Big\} = 
\mbox{Tr} \Big\{ |\Psi^{RS}\rangle \langle \Psi^{RS}|
\, (\openone^R \otimes {\cal E}^S) (|\Psi^{RS}\rangle \langle \Psi^{RS}|) 
\Big \} \:,
\label{fide1}
\end{equation}
{\it i.e.}, $F_e$ measures the fidelity between the input and output states 
of the joint system: $F_e=\mbox{Tr} \{ \rho^{RS} \rho^{RS'} \}$. $F_e$ can be 
expressed in terms of quantities intrinsic to the system alone once an 
operator-sum representation for ${\cal E}^S$ is available. If ${\cal E}^S(\rho^S)= 
\sum_\mu A_\mu^S \rho^S A_\mu^{S \,\dagger}$, Schumacher showed that 
\cite{schumacher}
\begin{equation}
F_e(\rho^S, {\cal E}^S ) = \sum_\mu \left| \mbox{Tr} \{ \rho^S A_{\mu}^S \}\right|^2 \:.
\label{fide}
\end{equation}

Because $F_e(\rho^S, {\cal E}^S)=1$ if and only if $\rho^{RS'}=|\Psi^{RS}
\rangle \langle \Psi^{RS}|$ \cite{schumacher}, $F_e$ naturally quantifies the 
{\sl preservation of quantum information} 
-- perfect preservation corresponding to implementing ${\cal E}^S=\openone^S$.
In the presence of the target transformation $U \equiv U^S \not = \openone^S$, 
the appropriate measure should equal 1 if and only if $\rho^{RS'}=
U^S |\Psi^{RS}\rangle \langle \Psi^{RS}| U^{S\,\dagger}$. Thus, (\ref{fide1}) is
generalized to a {\sl gate entanglement fidelity} as follows:
\begin{eqnarray*}
F_e (U^S \rho^S U^{S\,\dagger}, {\cal E}^S)= 
\mbox{Tr} \Big\{ U^S |\Psi^{RS}\rangle 
\langle \Psi^{RS}| U^{S\,\dagger}
\, (\openone^R \otimes {\cal E}^S) (|\Psi^{RS}\rangle \langle \Psi^{RS}|) 
\Big \} \:.
\end{eqnarray*}
By using the above operator-sum representation for ${\cal E}^S(\rho^S)$, 
one can derive the equivalent expression
\begin{eqnarray*}
F_e(U^S \rho^S U^{S\,\dagger}, {\cal E}^S)= 
\mbox{Tr} \left\{ |\Psi^{RS}\rangle \langle \Psi^{RS}|
\, (\openone^R \otimes \tilde{\cal E}^S) 
(|\Psi^{RS}\rangle \langle \Psi^{RS}|) \right\} = 
F_e( \rho^S, \tilde{\cal E}^S ) \:,
\end{eqnarray*}
where the modified dynamical map 
$\tilde{\cal E}^S = U^{S\,\dagger} {\cal E}^S U^S$ is defined by the set
of transformed Kraus operators $\{ U^{S\,\dagger} A_\mu^S \}$. 
Thus, a perfect implementation of the desired gate $U^S$ corresponds to 
perfect preservation of quantum information under $\tilde{\cal E}^S $:
the meaning of this is simply that, in the ideal case, the intended effect 
would be ${\cal E}^S (\rho^S) = U^S \rho^S U^{S\,\dagger}$, which is equivalent
to ensuring $\tilde{\cal E}^S (\rho^S) = \rho^S$.
Similar to (\ref{fide}), we then have
\begin{equation}
F_e( \rho^S, \tilde{\cal E}^S ) = \sum_\mu 
\left| \mbox{Tr}\{\rho^S U^{S\dagger} A_\mu^S \} \right|^2 \:,
\label{fide4}
\end{equation}
where as above $\rho^S= \mbox{Tr}_R|\Psi^{RS}\rangle \langle \Psi^{RS}|$ is 
the initial density matrix of the system alone.
Taking as the standard reference state a maximally entangled purification
$|\Psi^{RS}\rangle$ for which $\rho^S$ is the fully mixed state {\it i.e.},
$\rho^S=\openone^S/N$ for a $N$-dimensional state space, (\ref{fide4}) finally 
becomes
\begin{equation}
F_e(\tilde{\cal E}^S ) = \sum_\mu 
\left| \mbox{Tr} \{U^{S\dagger} A_{\mu}^S \}/N \right|^2 \:.
\label{fide5}
\end{equation}
This form makes it explicit that the gate fidelity defined in \cite{control} is
identical with the gate entanglement fidelity formally introduced here. 
We shall still refer to the quantity in (\ref{fide5}) simply as entanglement 
fidelity $F_e$ in the following.

The above reliability measure can be related to experimentally available data, 
and the results take particularly simple expressions in the case of single-qubit 
transformations we are concerned with. Starting from the 
standard maximally entangled Bell state for the joint $RS$ system, 
where $S$ and $R$ are now two qubits, and assuming that the process ${\cal E}^S
\equiv {\cal E}$ actually implementing $U$ 
is unital and trace-preserving, one finds 
\begin{equation} 
F_e = {1 \over 2} \Big( F_{U |0\rangle} + F_{U |+ \rangle} + F_{U |+i\rangle} 
-1 \Big) \:, 
\label{finalfid}
\end{equation}
where $F_{U |\psi_{in} \rangle} = \mbox{Tr} \{ U |\psi_{in}\rangle \langle \psi_{in}| 
U^\dagger {\cal E}(|\psi_{in}\rangle \langle \psi_{in}|) \}$ for a generic one-bit pure 
input $|\psi_{in}\rangle$, and $|0\rangle$, $|+\rangle$, $|+i\rangle$ are eigenstates 
with positive eigenvalue of $\sigma_z, \sigma_x, \sigma_y$, respectively. 
This expression was used in \cite{ns} for the special case $U=\openone$.
As a further remark, it is worth noting that $F_e$ as given in (\ref{finalfid})
is related to the so-called {\sl average gate fidelity} $\overline{F}$ proposed in 
\cite{jones} via $\bar{F}= 2/3 \,F_e + 1/3$.  Eq. (\ref{finalfid}) is directly 
applicable to quantifying the accuracy of DF unitary manipulations as given, upon 
decoding, by (\ref{out}).

\section{Demonstration of a decoherence-free qubit}

The utility of a DFS to preserve quantum information ({\it i.e.}, to implement 
the identity operation) is demonstrated under the action of different classes of 
both engineered and natural noise: a variable-strength engineered 
decoherent noise, a full-strength (crusher) engineered incoherent noise, and the 
natural ambient noise due to relaxation.  A significant improvement in the 
entanglement fidelity for each class of noise is seen.  In addition, because
the measured entanglement fidelities remain above the threshold value 0.50 
\cite{bennett}, all implementations guarantee, in principle, the ability to 
preserve entanglement over a wide range of noise strengths. 
Unlike the case of an NS, the entire state of the system (data plus ancilla) 
remains unchanged under the action of the noise.  While this was experimentally 
confirmed to a good accuracy \cite{DFSNSComment}, we report the preservation of 
quantum information between the desired input state and the measured output 
state of the data qubit alone.

\subsection{Engineered noise} 

Gradient-diffusion techniques were used to implement variable-strength noise.
In order to isolate the effects of the applied noise, the time delay between 
encoding and decoding was kept fixed.  In the first half of this time delay
collective phase noise was applied to the system.  Unwanted evolution due to 
the internal Hamiltonian was refoucsed during the second half of the delay 
by a pair of $\pi$ pulses ${\cal P}_2$ given in Sect. IIIa.  The gradient 
strength was varied over the full dynamic range of the spectrometer (0 to 60 
Gauss/cm) and the diffusion time $\Delta$ was set in such a way that a 
significant amount of information was lost when the un-encoded data spin was 
directly exposed to noise. This was obtained by running separate experiments
with encoding/decoding sequences turned off. 

The experimental data are collected in Fig. 2.
For the encoded data, assuming no additional loss of information with increasing 
noise strengths is consistent with the experimental results.  Fitting the data 
with a constant value yields a value of $0.97 \pm 0.01$.  
The deviation from unity is consistent with the observed $F_e$ value for the 
reference situation of no applied noise ({\it i.e.}, the one 
implementing a net identity evolution between encoding and decoding).  
Therefore, these losses are caused by imperfections in the applied pulses
as well as by natural noise processes whose action is not in the correctable 
error algebra ${\cal A}_z$ (see below).  For all but the smallest noise values, a
substantially increased amount of quantum information is retained using the DFS 
memory than leaving the system un-protected.

To further confirm the robustness of the DFS memory against arbitrary noise
strengths, an incoherent implementation of all possible collective phase errors 
was also realized to emulate noise in the strong dephasing limit.  A single
magnetic field gradient pulse with maximum strength was applied, causing spins 
on the fringe of the sample to evolve through more than 750 cycles and 
therefore acquire large phase errors. 
Again, the loss of information in the presence of this crusher noise is 
compatible with the measured loss due to just encoding and decoding 
(see Table 1).

\subsection{Natural noise}

The behavior of both the DFS-encoded and the un-encoded data under ambient 
noise was also probed in a separate series of experiments, with the goal of 
gaining qualitative insight on the relevance of fully correlated dephasing 
in the naturally occurring phase relaxation processes.  In this case, the 
holding time between encoding and decoding was varied to allow for a variable 
exposure to noise.  Because natural relaxation takes significant contributions 
from $T_1$ processes (that are both amplitude and phase damping) in addition 
to transverse $T_2$ relaxation, the unitality assumption invoked in deriving 
the expression (\ref{finalfid}) for the entanglement fidelity is no longer 
accurate. While $F_e$ could be still evaluated directly from (\ref{fide5}) upon 
experimentally extracting a set of Kraus operators, a simpler coherence metric 
is appropriate if the actual amplitude decay is of no interest.  
Similar to \cite{kielpinski}, the amount of quantum coherence ${\cal C}$ (phase 
information) that is retained in the course of the noisy evolution can be quantified
by experimentally determining the average off-diagonal component present in the output 
density matrix. Thus, corresponding to the two experimentally prepared transverse 
p.p. states $|+\rangle, |i\rangle$ defined above, we calculate
\begin{eqnarray*}
{\cal C} = {1 \over 2} \Big(   
\mbox{Tr} \{ \sigma_x {\cal E}( |+\rangle \langle +|) \} +
\mbox{Tr} \{ \sigma_y {\cal E}( |i\rangle \langle i|) \} \Big) \:,
\end{eqnarray*}
where the map ${\cal E}$ now corresponds to the natural noisy dynamics. 

The experimental data are presented in Fig. 3.  Holding times ranging from 
a fraction of a second up to a time scale comparable to $T_2$ were explored. 
An appreciable decay of the DFS qubit is seen in this case, witnessing the 
presence of non-collective phase-damping processes in the ambient noise. 
In spite of this non-robust behavior, the DFS is still able to retain quantum 
coherence much longer than the un-encoded state. This implies that a significant 
contribution to the overall phase relaxation is actually caused by fully 
correlated dephasing, consistent with the physical intuition based on the 
geometrical and chemical structure of the molecule.

\section{Demonstration of universal control over a decoherence-free qubit}

As stated in section IIc, evolution according to two non-commuting encoded 
Hamiltonians is required for universal control. In the implementation, 
we found it convenient to adopt a choice of encoded observables intermediate 
between (\ref{obs1}), (\ref{obs2}), {\it i.e.}
$\sigma_z^L=-\sigma^2_z$, $\sigma_x^L=(\sigma_x^1\sigma_x^2+\sigma_y^1
\sigma_y^2)/2$ henceforth. In terms of this choice, and using the actual 
implementation parameters (see Fig. 1), the internal Hamiltonian (\ref{hint3}) 
is given by
\begin{equation}
H_{int}=_L -137.5\pi\sigma_z^{\it L} + 5.7\pi\sigma_x^{\it L}\:.
\label{hintexp}
\end{equation}

\subsection{Encoded $z$ and $x$ rotations}

According to the above expression, the strong $\sigma_z^L$ Hamiltonian 
dominates the $\sigma_x^L$ contribution. 
Therefore, an encoded $z$ evolution can be implemented, to high accuracy, 
by simply waiting the appropriate amount of time.  Because of the negative 
sign in (\ref{hintexp}), a positive $\sigma_z^L$ rotation of $\theta$ can 
be implemented by rotating by $2\pi-\theta$ about the negative logical 
$z$ axis.  A representative $\pi/2$ encoded rotation, $\exp(-i\pi\sigma_z^L/4)$, 
was experimentally implemented with a fidelity of entanglement of $0.94 \pm 0.03$.  
Comparing this result with the accuracy of the identity operation (see Table 1), 
we see no significant loss of information due to the $z$ gate.

Implementing a $x$ rotation is less straight forward. 
While a logical $x$ Hamiltonian is present in (\ref{hintexp}), 
it is quickly averaged out by the stronger logical $z$ term.  
With the encoded system expressed in this form, it is clear
that we must average out the $\sigma_z^L$ term at a rate faster than
its strength \cite{ernst}, using control operations that commute with 
$\sigma_x^L$.
The external RF Hamiltonian provides this extra control parameter, at the 
expenses of forcing the system to leave the DFS -- if only for short periods 
of time.  While continuous irradiation of the spins would suffice to preserve 
the $\sigma_x^L$ Hamiltonian by matching the so-called Hartman-Hahn condition 
\cite{HH,tocsy}, as discussed in Sect. IIIb a train of $\pi^L$ pulses achieves 
the same goal \cite{CP,Hahn,MG}, with the advantage that the system may be 
left in the DFS for significant portions of the control time.  
The trick is to note that, although the evolution induced by the applied RF 
field does not resemble in general a $\sigma_x^L$ Hamiltonian, for the 
special case of a hard $\pi$ pulse the resulting propagator is
\begin{eqnarray*}
U_h^{1,2}=\exp\Big(\hspace*{-1mm}-i\frac{\pi}{2}(\sigma_x^1+\sigma_x^2)\Big)=
\exp\Big(\hspace*{-1mm}-i\frac{\pi}{2}\sigma_x^1\Big)
\exp\Big(\hspace*{-1mm}-i\frac{\pi}{2}\sigma_x^2\Big)=
-\sigma_x^1\sigma_x^2 \:,
\end{eqnarray*}
which mimics a net $\sigma_x^L$ operation: the action of $U_h^{1,2}$ 
on the code subspace is identical to the action of $\sigma_x^L$ {\sl as a unitary 
operator} (not as a Hamiltonian -- note that $\sigma_x^1+\sigma_x^2$ does not 
clearly respect the form (\ref{goodham})).

Composite pulses \cite{Freeman}, which provide an excellent balance 
between speed and robustness, were used to implement each hard $\pi$ pulse.  
Six-period pulses optimized to be robust against variations in both chemical 
shift (phase errors) and RF strength (control errors) \cite{Shaka1} were used 
to emulate the required sequence of encoded $\pi^L$.  Each of these hard pulses 
is $62.4 \mu$s in duration and is followed by a delay of $630 \mu$s.  Therefore, 
the system resides in the protected space for over $90\%$ of the computational 
time.  The phases of the $\pi^L$ pulses were alternated systematically as described 
by a WALTZ sequence \cite{Shaka2}, so as to minimize the impact of experimental 
errors.  In particular, a 64-cycle sequence was used to achieve a $\pi/2$ encoded 
rotation, $\exp(-i \pi \sigma_x^L/4)$, with a fidelity of entanglement of 
$0.94 \pm 0.03$.  Again, we see no significant loss of information due to the $x$ 
operation.

\subsection{Composite encoded $y$ rotation under collective phase noise}

To explicitly test the robustness of the available logical $x$ and $z$ manipulations, 
a composite encoded rotation by $\pi/2$ about $y$ was implemented in the presence 
of variable-strength collective phase noise {\it i.e.}, the sequence of 
encoded rotations
\begin{eqnarray*}
\exp{\Big(\hspace*{-1mm}-i {\pi \over 4} \sigma_z^L\Big)}  
\exp{\Big(\hspace*{-1mm}-i {\pi \over 4} \sigma_x^L\Big)}  
\exp{\Big(i {\pi \over 4} \sigma_z^L\Big)} =
\exp{\Big(\hspace*{-1mm}-i {\pi \over 4} \sigma_y^L\Big)}   
\end{eqnarray*}
was performed.
As in the DFS memory experiments, gradients were used to induce a spatially 
incoherent error over different parts of the sample. However, the noise 
effects associated with a time-independent gradient Hamiltonian ({\it i.e.,} 
an infinitely long correlation time) would tend to be effectively averaged out 
over time by the applied control sequences as described by coherent averaging 
\cite{Waugh} and dynamical decoupling \cite{viola3}. 
In order to make sure that the net action of the applied gradients is maintained
in the presence of the external control, a procedure similar to the 
{\sl fast-switching} control schemes discussed in \cite{viola4} was followed,  
by rapidly modulating the strength of the applied gradient Hamiltonians over 
the course of the control sequence. Thus, a temporal incoherence was also 
superimposed at each spatial location in the sample, enforcing a finite
correlation time $\tau_c$ (hence a non-zero cut-off frequency) in the spectral 
density describing the noise. In practice, the gradient waveform was determined 
via a random walk process, whose shape is depicted in Fig. 4. The gradient
strength was changed every 50.6 $\mu$s therefore $\tau_c \sim 50.6 \mu$s.
By making sure that $\tau_c$ is short compared to the control cycle time of
the sequences used to implement the composite rotations ($\sim 700 \, \mu$s
in our case), active control is made ineffective at averaging out the high
frequency effects of the noise during the computation.   

A broad range of values for the maximum applied gradient strength were explored
to test the robustness of the computation to collective phase errors. 
The experimentally determined gate entanglement fidelities are shown in Fig. 5.  
As expected, the computation is protected up to a particularly noise intensity 
and then falls off with increasing noise strength.  As in the memory case, it is 
worth stressing that $F_e$ values well exceeding the value 0.50 have been achieved 
over the entire range of applied noise strengths.  It should be noted that 
because the active sample is order $1$ cm, most of the sample is experiencing noise 
strengths significantly stronger than natural fluctuations -- which are approximately 
1 Hz in strength.

\section{Conclusions}

We have provided the first demonstration of universal control over a DFS-encoded 
qubit. The implementation relied on combining the benefits of passive noise 
protection via DFS coding with the ability of relaxing the constraints of fully 
DF manipulations via appropriate control design --
thereby also validating the underlying principles of encoded decoupling.
We believe that our techniques are applicable to a wide class of
quantum information devices, where collective dephasing mechanisms 
play a dominant role and where the structure of the system's internal 
Hamiltonian can be mapped onto a NMR-type Hamiltonian. These may include
various solid-state proposals as discussed in \cite{exchange,wu}.  
Thus, our results improve the prospects that DFS/NS coding, combined with 
encoded dynamical decoupling and robust control design, will play a practical 
role for both protected storage and manipulation of quantum information in QIP.

\section{Acknowledgements}
This work was supported by the National Security Agency and Advanced 
Research and Development Activity under Army Research Office contract number 
DAAD19-01-1-0519, by the Defense Sciences Office of the Defense Advanced 
Research Projects Agency under contract number MDA972-01-1-0003, and by 
the Department of Energy under contract W-7405-ENG-36.  It is a pleasure to 
thank Chandrasekhar Ramanathan, Joseph Emerson, and Tim Havel for useful 
discussions. 


\vspace*{10mm}

\begin{table}
\begin{tabular}{ccccc}
&   &   &   \vspace*{-3mm}   &  \\
$\;\;${\sf Quantum process}$\;\;\;\;$ & $\;\;\;F_{|0\rangle}\;\;\;$
                      & $\;\;\;F_{|+\rangle}\;\;\;$
                      & $\;\;\;F_{|+i\rangle}\vspace*{1mm} \;\;\;$
                      & $\;\;\;\; {\bf F_e}\;\;\;\;$ \\ \hline\hline
                  &   &   &   \vspace*{-3.5mm}   &  \\
  ${\cal Q}_{z,{\sf un}}$  & 1.00 & 0.50 & 0.50 & {\bf 0.50} \\ \hline
                 &   &   &   \vspace*{-3.5mm}   &  \\
  ${\cal Q}_{0,{\sf df}}$  & 0.98 & 0.95 & 0.94 & {\sf 0.93} \\
  ${\cal Q}_{z,{\sf df}}$  & 0.97 & 0.98 & 0.96 & {\sf 0.95} \\ 
\end{tabular}
\vspace*{1mm}
\caption{Experimental data for the implementation
of full-strength collective dephasing. Input-output fidelities and entanglement 
fidelities corresponding to the application of the intended error model to 
both the DFS encoded (${\cal Q}_{z,{\sf df}}$) and the un-encoded data spin 
(${\cal Q}_{z,{\sf un}}$) are listed, along with the values relative to the 
reference situation of zero applied noise between DFS encoding and decoding
(${\cal Q}_{0,{\sf df}}$). Crusher gradient fields with full strength 
$\sim 60$ Gauss/cm were applied for a period $\delta = 745 \mu s$. 
The measured values for the un-encoded test data confirm the expectation that 
the applied noise process induces full phase damping on the data spin, with 
predicted $F_e=0.50$. Systematic uncertainties are $\sim 0.02$ while statistical 
uncertainties are $\sim 2\%$, both due to errors in the tomographic density 
matrix reconstruction. } 
\end{table}

\begin{figure}[h!]
\begin{center}
{\epsfysize=3.0in\centerline{\epsffile{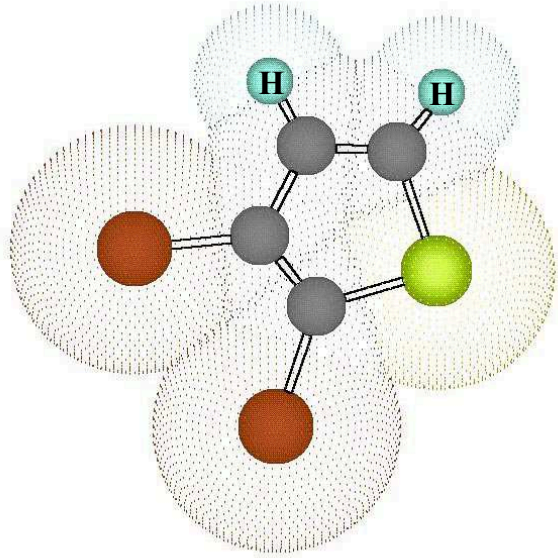}}}
\end{center}
\caption{Molecular structure of Dibromothiophene. The two proton qubits are 
indicated. As spectroscopically the two protons are effectively undistinguishable, 
the qubit labels have a purely formal meaning. 
All experiments were carried out in a magnetic field of $\sim 9.7$ 
T with one proton on resonance.  The frequency shifts of the second 
proton is $\nu_2=137.5$ Hz, while the $J$-coupling constant is $J=5.7$ Hz. 
The longitudinal and transverse relaxation times are $T_1 \sim 7$ s and 
$T_2 \sim 3.5$ s, respectively. } 
\end{figure}

\begin{figure}
\begin{center}
{\epsfysize=3.0in\centerline{\epsffile{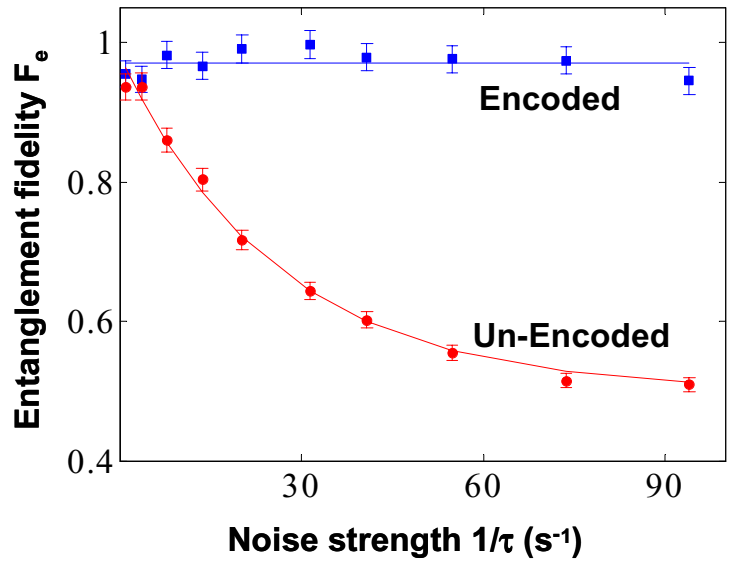}}}
\end{center}
\caption{ Experimentally determined entanglement fidelity for the implementation
of variable-strength collective dephasing. Both the behavior of the DFS-encoded 
(squares) and the un-encoded (circles) data is shown. The independent axis (noise 
strength) was determined by fitting the un-encoded data to an exponential decay 
of the form $F_e = A \exp({-t_{ev}/\tau})+0.5$, with $t_{ev}=
\Delta+2\delta=37.765$ ms.  The un-encoded data is only displayed for reference.  
The encoded data is fit to a constant value $F_e=C$, yielding the 
best estimate $C=0.97 \pm 0.01$.  Systematic uncertainties (not included in
the figure) are $\sim 0.02$.}
\end{figure}

\begin{figure}
\begin{center}
{\epsfysize=3.0in\centerline{\epsffile{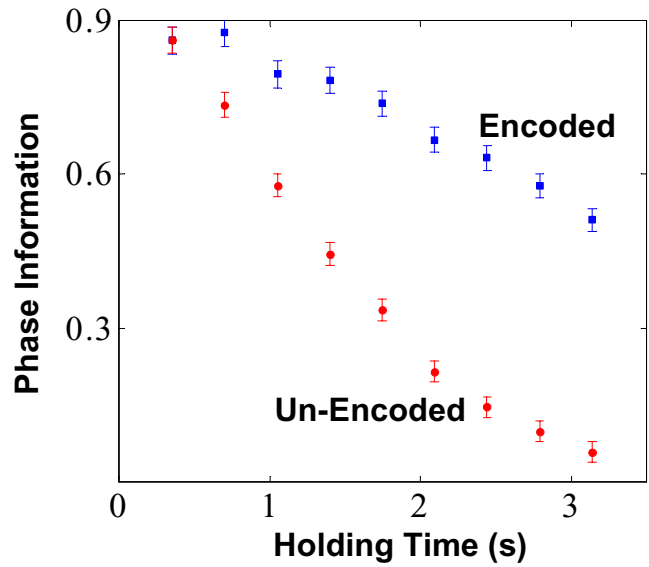}}}
\end{center}
\caption{Experimental data for the phase information retained after exposure to 
the natural system noise.  The average preservation of $\sigma_x$ and $\sigma_y$ 
was measured as a function of holding times from 0 to $\sim 3$ s.  
Improvement over the un-encoded case is seen, confirming that collective phase 
errors are one of the dominating modes of natural noise for this system. } 
\end{figure}

\begin{figure}
\begin{center}
{\epsfysize=3.0in\centerline{\epsffile{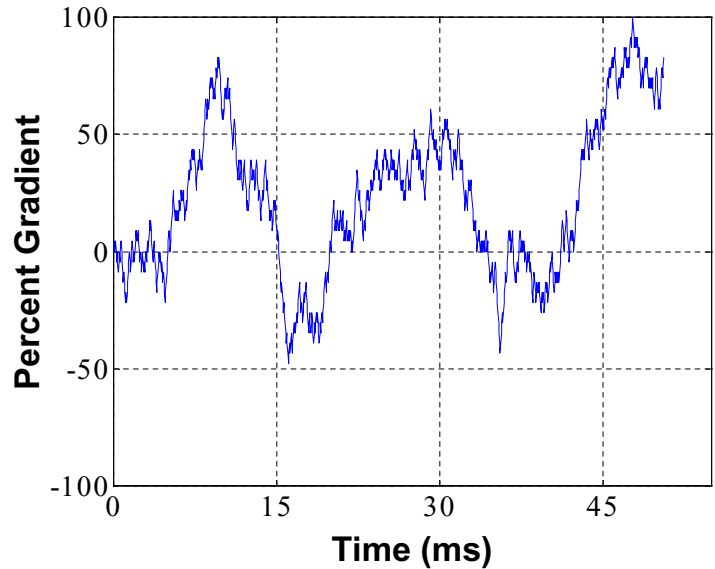}}}
\end{center}
\caption{The temporal variation of the gradient waveform used to test 
control of a DFS qubit in the presences of noise.  The shape was determined 
by a random walk algorithm.  The stepping time was $50.6\mu s$, which is faster
than any of the control time scales relevant in implementing the encoded 
transformations. } 
\end{figure}

\begin{figure}
\begin{center}
{\epsfysize=3.0in\centerline{\epsffile{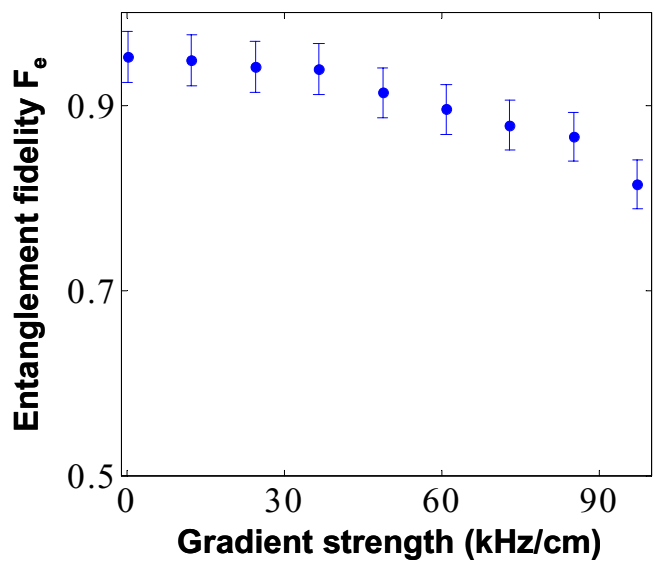}}}
\end{center}
\caption{Experimentally determined entanglement fidelity for the implementation
of a composite encoded $y$ rotation of $\pi/2$ in the presence of noise.  
Magnetic field gradients implement a spatially incoherent collective phase 
error as a function of the molecular position.  Gradient strengths from 0 to 
$\sim 100$ kHz/cm were applied over a 1 cm sample. The behavior of $F_e$ 
is flat over a broad range of noise strengths and remains significantly above 
the $0.5$ threshold for all noise values considered.  This convincingly demonstrates 
the ability to control a DFS qubit in the presence of noise significantly stronger 
than the natural noise of the system.} 
\end{figure}

\end{document}